\documentclass[12pt]{article}

\newenvironment{wileykeywords}{\textsf{Keywords:}\hspace{\stretch{1}}}{\hspace{\stretch{1}}\rule{1ex}{1ex}}

\usepackage{amsmath,amssymb}
\usepackage{graphicx}
\usepackage{color}
\usepackage{dcolumn}
\usepackage{bm}
\usepackage[numbers,super,comma,sort&compress]{natbib}

\definecolor{background-color}{gray}{0.98}

\title{The Behavior of Benzene Confined in Single Wall Carbon Nanotube}
\author{Yu. D. Fomin(1,2), E.N. Tsiok(1) and V.N. Ryzhov(1,2) \thanks{(1)Institute for High Pressure Physics Russian Academy of Sciences,
(2) Moscow Institute of Physics and Technology (State
University)}}

\begin{document}

\maketitle

\begin{abstract}
We present the molecular dynamics study of benzene molecules
confined into the single wall carbon nanotube. The local structure
and  orientational ordering of benzene molecules are investigated.
It is found that the molecules mostly group in the middle distance
from the axe of the tube to the wall. The molecules located in the
vicinity of the wall demonstrate some deviation from planar shape.
There is a tilted orientational ordering of the molecules which
depends on the location of the molecule. It is shown that the
diffusion coefficient of the benzene molecules is very small at
the conditions we report here.
\end{abstract}

\begin{wileykeywords}
Benzene, carbon nanotubes, orientational order, structure.
\end{wileykeywords}

\clearpage



  \makeatletter
  \renewcommand\@biblabel[1]{#1.}
  \makeatother

\bibliographystyle{apsrev}

\renewcommand{\baselinestretch}{1.5}
\normalsize

\clearpage

\section*{\sffamily \Large SUMMARY}


It is well known that confining a liquid into a pore strongly
alters the liquid behavior. Investigations of the effect of
confinement are of great importance for many scientific and
technological applications. Here we present a study of the
behavior of benzene molecules confined in the single wall carbon
nanotubes. We find that the molecules of benzene mostly group in
the middle distance from the axe of the tube to the wall. The
molecules located in the vicinity of the wall demonstrate some
deviation from planar shape. There is no any strong ordering in
the system, however, there is a tilted orientation of molecules
which depends on the location of the molecule.

\section*{\sffamily \Large INTRODUCTION} 

It is well known that materials confined in nanoscale dimensions
have properties that strongly differ from the properties of bulk
systems. This is due to the reducing the dimensionality of the
system and interface effects. Confining boundaries bias the
spatial distribution of the constituent molecules and the ways by
which those molecules can dynamically rearrange. These effects
play important roles in the thermodynamics of the confined systems
and influence the topology of the phase diagram \cite{rev1,rice}.
The confinement can drastically change the thermodynamic
parameters of phase transitions and even form the new phases due
to interaction with the boundaries. For example, the melting
temperature of confined benzene depends on the form and size of
the nanopores. In most cases the melting point of the solid in the
pore decreases with decreasing pore diameter \cite{rev1}. In Ref.
\cite{ben1} the melting behavior of the confined benzene was
discussed for different types of porous confinement. It was shown
that no crystallization is observed in the cylindrical pores below
the pore size about $4.7 nm$. This value corresponds approximately
to 10 molecular diameters. It was shown that benzene does
crystallize in $4.7 nm$ pores but vitrifies in narrower pores
\cite{ben2}. On the other hand, as it was shown in Refs.
\cite{ch1,ch2,ch3,ch4}, experimental freezing of
tetrachloromethane was observed in activated carbon fibers for the
pore widths up to $0.75 nm$ (less than two molecular diameters),
whereas cylindrical pores of several molecular sizes are necessary
to have crystallization of tetrachloromethane molecules.

The general motivation for the study of different nanoconfined
systems follows from the fact that there are a lot of real
physical and biological phenomena and processes that depend on the
properties of such systems and play an important role in the
different fields of modern technology. However, nanoconfinement is
considerably interesting also due to the new physics observed in
these systems. For example, fluids confined in carbon nanotube
exhibit formation of layers, crystallization of the contact layer
\cite{cryst} and a superflow which depends on the confinement
\cite{d0,d1,d2,d3,d4}.

In this paper, we present a systematic molecular dynamics study of
single wall carbon nanotubes (SWCN) \cite{cn1,cn2} doped with
benzene ($C_6H_6$) molecules. Carbon nanotubes are widely
investigated first of all because of their potential applications
in material science, biotechnology and medicine
\cite{app1,app2,app3,app4}. Introducing molecules into carbon
nanotubes can drastically change the electronic properties of
nanotubes. Benzene molecules are widely used for study the
influence of doping on the properties of carbon nanotubes because
of their small size which permits them to be encapsulated easily
inside carbon nanotubes of different diameters. The orientation
and the position of the benzene molecules inside the carbon
nanotubes determine the optical, magnetic and electrical transport
properties of the whole system \cite{be1,be2,be3}.

In Ref. \cite{tworings} a semi-analytical model for the
interaction of a benzene molecule and a carbon nanotube was
proposed. It was shown that the orientation of the molecule
drastically depends on the radius of the nanotube. The authors
found that horizontal, tilted and perpendicular equilibrium
configurations are possible for the benzene molecule on the axis
of the carbon nanotube when the radius of it is less than $5.580
\AA$. However, when the radius of the nanotube is larger than
$5.580 \AA$, the equilibrium configurations occur at an offset
horizontal orientation.

The results of Ref. \cite{tworings} were obtained for one
molecule. The main goal of this article is to study the
positional, orientational and dynamic behavior of the ensemble of
benzene molecules in the single wall carbon nanotube.

\section*{\sffamily \Large METHODOLOGY}




In the present article we study benzene molecules in carbon single
wall nanotube by means of the molecular dynamics simulation. The
benzene molecule consists of a ring of $6$ carbon atoms. Each
carbon is also bonded with a hydrogen atom out of the ring. The
radius of benzene ring is about $2.8 \AA$. If one takes into
account the length of the hydrogen bonds then the size of benzene
molecule is approximately $5 \AA$. The radius of the nanotube is
$6.89 \AA$ and the length  is equal to $250.2363 \AA$. It means
that the radius of the nanotube is just a bit larger then the size
of the molecule. We choose this radius because it corresponds to
the situation when, as proposed in Ref. \cite{tworings}, one can
expect to find the offset horizontal orientation of the benzene
molecules.  The tube is oriented parallel to the $z$ axe. The
system has periodic boundary conditions in $z$ direction while it
is confined in $x$ and $y$ ones. Three systems were studied: $150,
200$ and $250$ benzene molecules in the nanotube described above.
The later system corresponds to the density $0.877 g/cm^3$ which
is the density of bulk benzene at ambient conditions. All systems
were simulated at three temperatures: $300K$, $400K$ and $500K$.
Doing this way we could check the influence of both density and
temperature on the behavior of benzene in nanotubes.

We simulate the system in canonical ensemble (constant number of
particles $N$, volume $V$ and temperature $T$). The temperature is
kept constant by applying Nose-Hoover thermostat. The carbon atoms
of the nanotube are held rigid in order to stabilize the system
while the benzene molecules are moved in molecular dynamics runs.
The timestep is $0.1 fs$. So small timestep was necessary in order
to correctly take into account the motion of hydrogens in the
benzene molecules.

All interactions in the system were described by AIREBO
interatomic potential \cite{airebo}. This potential is specially
developed for simulation of the systems containing carbon and
hydrogen atoms, and it allows to consider all interactions in the
system in framework of the same model.

All simulations were made in lammps simulation package
\cite{lammps}.

\section*{\sffamily \Large RESULTS AND DISCUSSIONS}



\begin{figure}
\includegraphics[width=15cm, height=5cm]{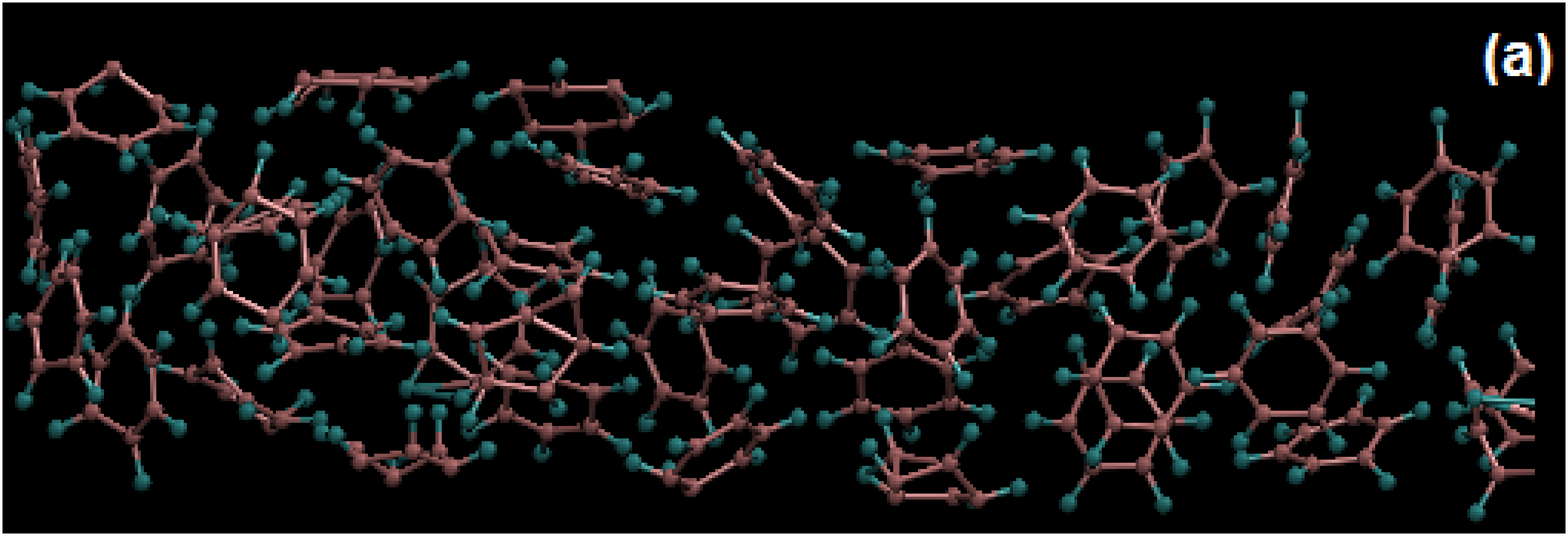}%

\includegraphics[width=7cm, height=7cm]{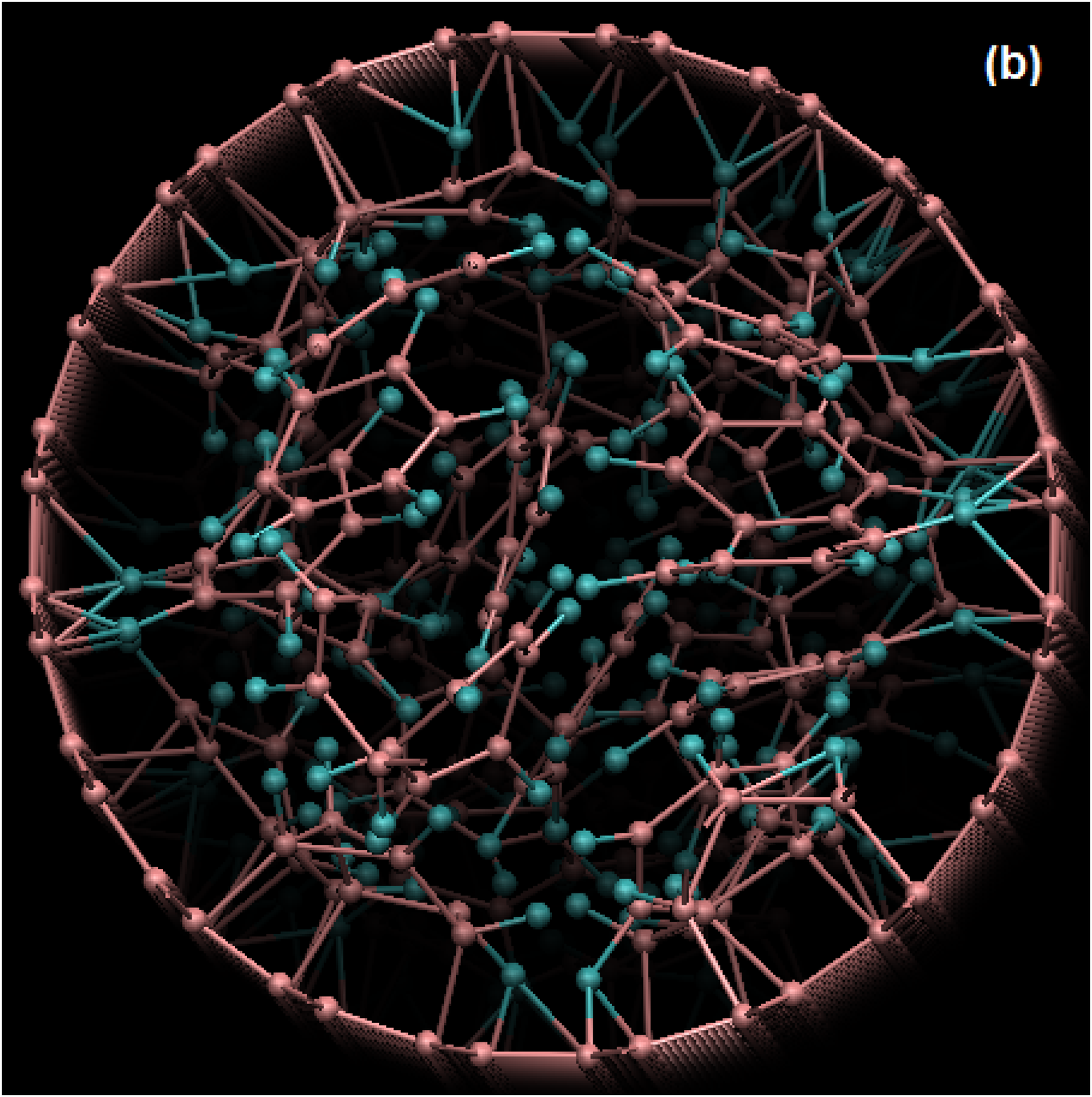}%

\caption{\label{fig:fig1} (a) Snapshot of benzene molecules in the
SWCN. The nanotube itself is not shown. (b) Cross section of the
system. $N=250$, $T=300 K$.}
\end{figure}

The qualitative ideas on the structure of benzene inside nanotube
can be obtained from the snapshots on the system.
Fig.~\ref{fig:fig1} (a) shows a part of the nanotube. In order to
make the view of the molecules clearer we do not show the tube
itself. One can see that the system does not demonstrate
pronounced order. Another important conclusion from the snapshots
is that if one looks at a cross section of the tube
(Fig.~\ref{fig:fig1} (b)) one can see that the centers of mass of
the molecule do not approach both the central axe and the walls
and prefer to be somewhere in the middle. In Ref. \cite{tworings}
a simple model of benzene in carbon nanotube was proposed. Basing
on this model, the authors studied possible locations and
orientations of benzene molecules inside the nanotube. It was
found that for the tube radius $R<5.580 \AA$ the equilibrium
position of the molecule belongs to the central axe while for
higher radiuses the position moves apart. Our calculations are
made at $R=6.89 \AA$ in order to compare our results with this
publication.

\begin{figure}
\includegraphics[width=7cm, height=7cm]{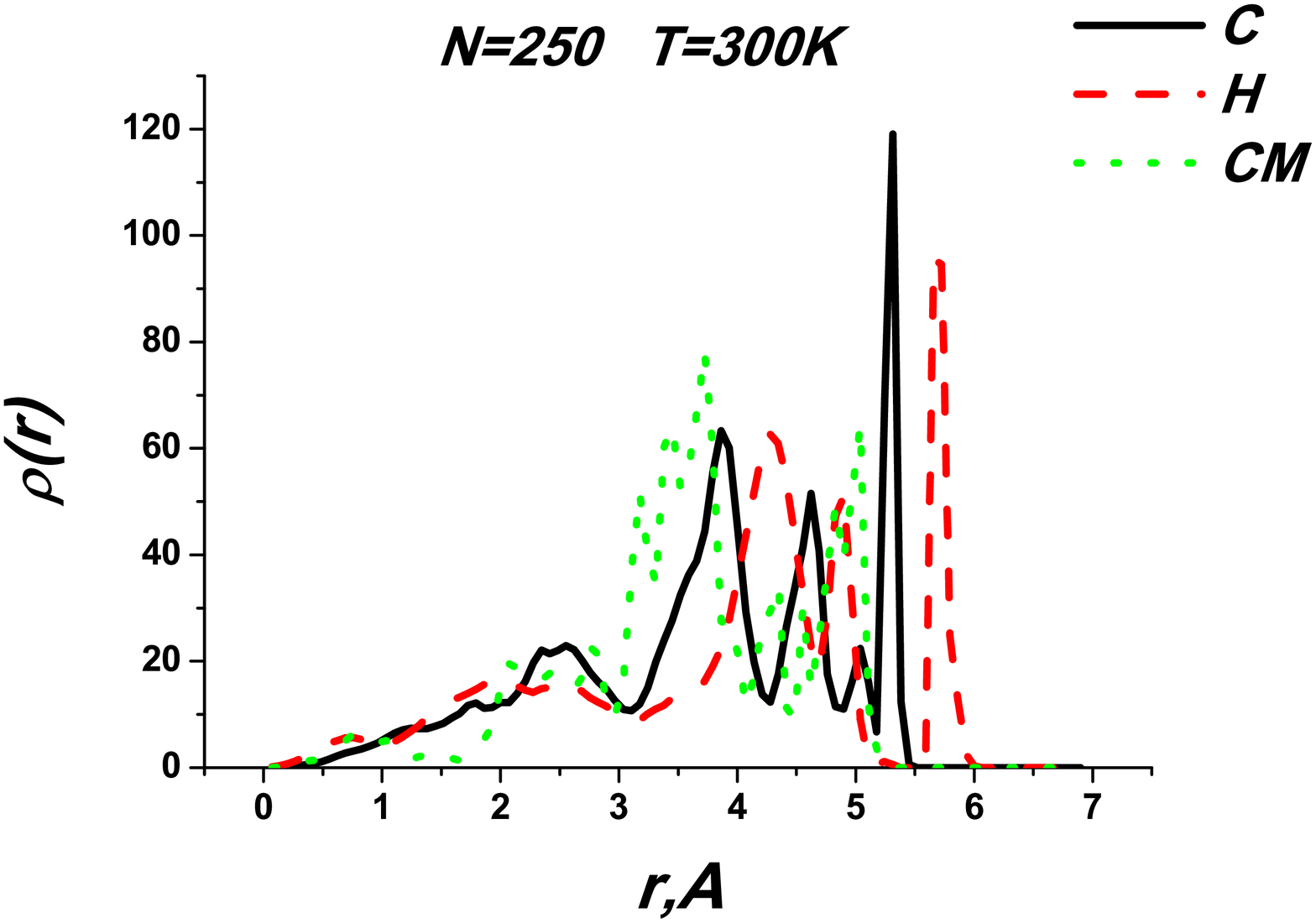}%

\caption{\label{fig:fig2} Radial distribution of number densities
of carbon, hydrogen and centers of mass of the molecules for
$N=250$ and $T=300K$.}
\end{figure}

The described features can be easily seen from the density
profiles of different species. Fig.~\ref{fig:fig2} shows the
profiles of number density of carbon and hydrogen species and of
the centers of mass of the molecules. The largest peak of the
carbon density distribution corresponds to the distance $5.3 \AA$
from the central axe of the tube. After that the curve rapidly
vanish. So the closest approach of carbon atoms to the wall of the
tube is approximately $1.5 \AA$.

However, the main peak of the centers of mass of the molecules is
closer to the center (Fig.~\ref{fig:fig2}). It corresponds to the
distance $r_{cm1}=3.73 \AA$. The second peak of the centers of
mass density is almost of the same height and located at
$r_{cm2}=5.04 \AA$.

One can see that the closest peak to the wall is the one of
hydrogen distribution. The maximum is located at $5.7 \AA$, i.e.
approximately $1 \AA$ from the wall. So close approach of
hydrogens to the nanotube can mean that some kind of effective
hydrogen bonds appear in the system.

Importantly, all distributions vanish at $r=0$ which means that at
this density the particles try to avoid the central axe of the
tube.

\begin{figure}
\includegraphics[width=7cm, height=7cm]{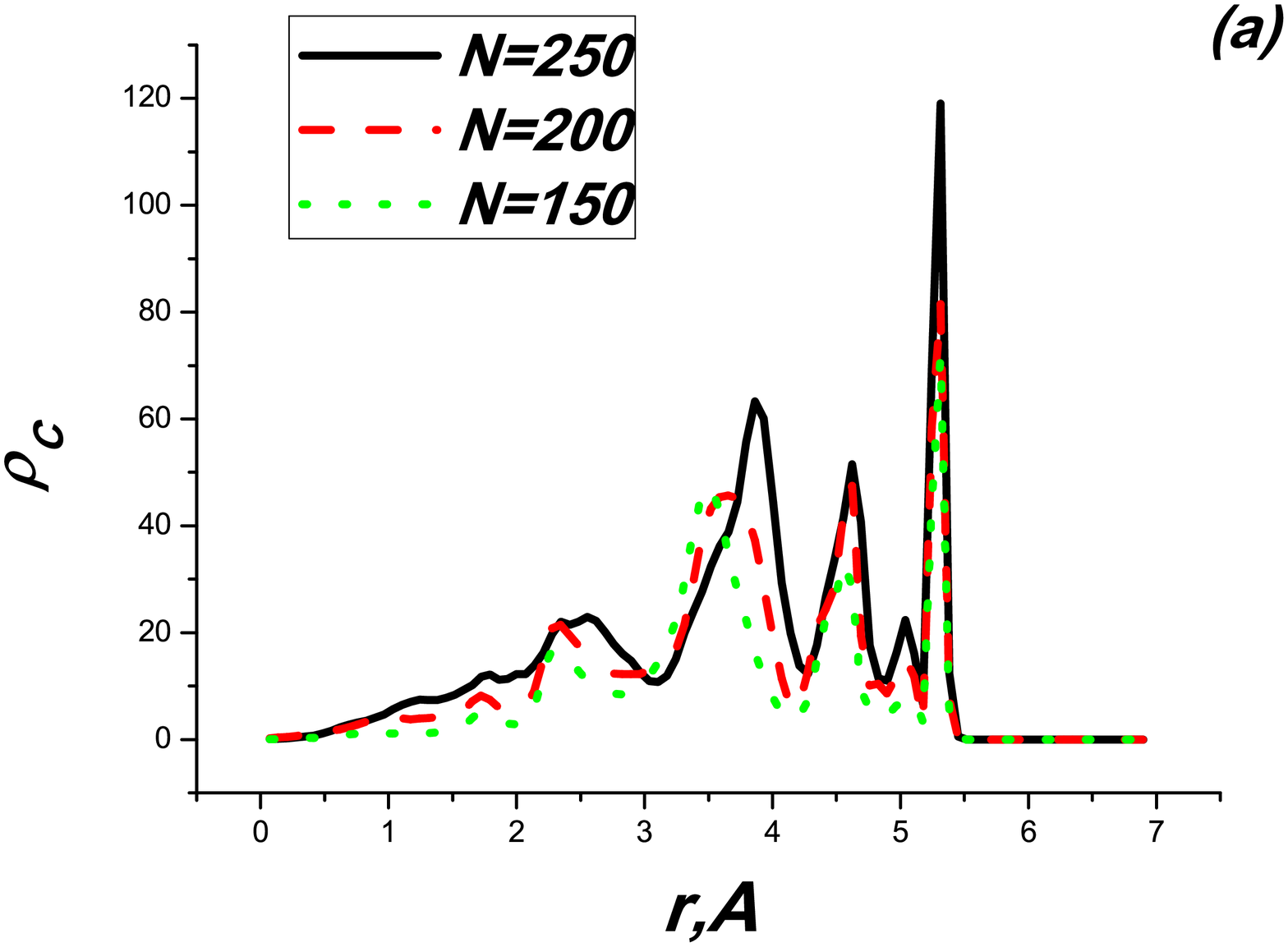}%
\includegraphics[width=7cm, height=7cm]{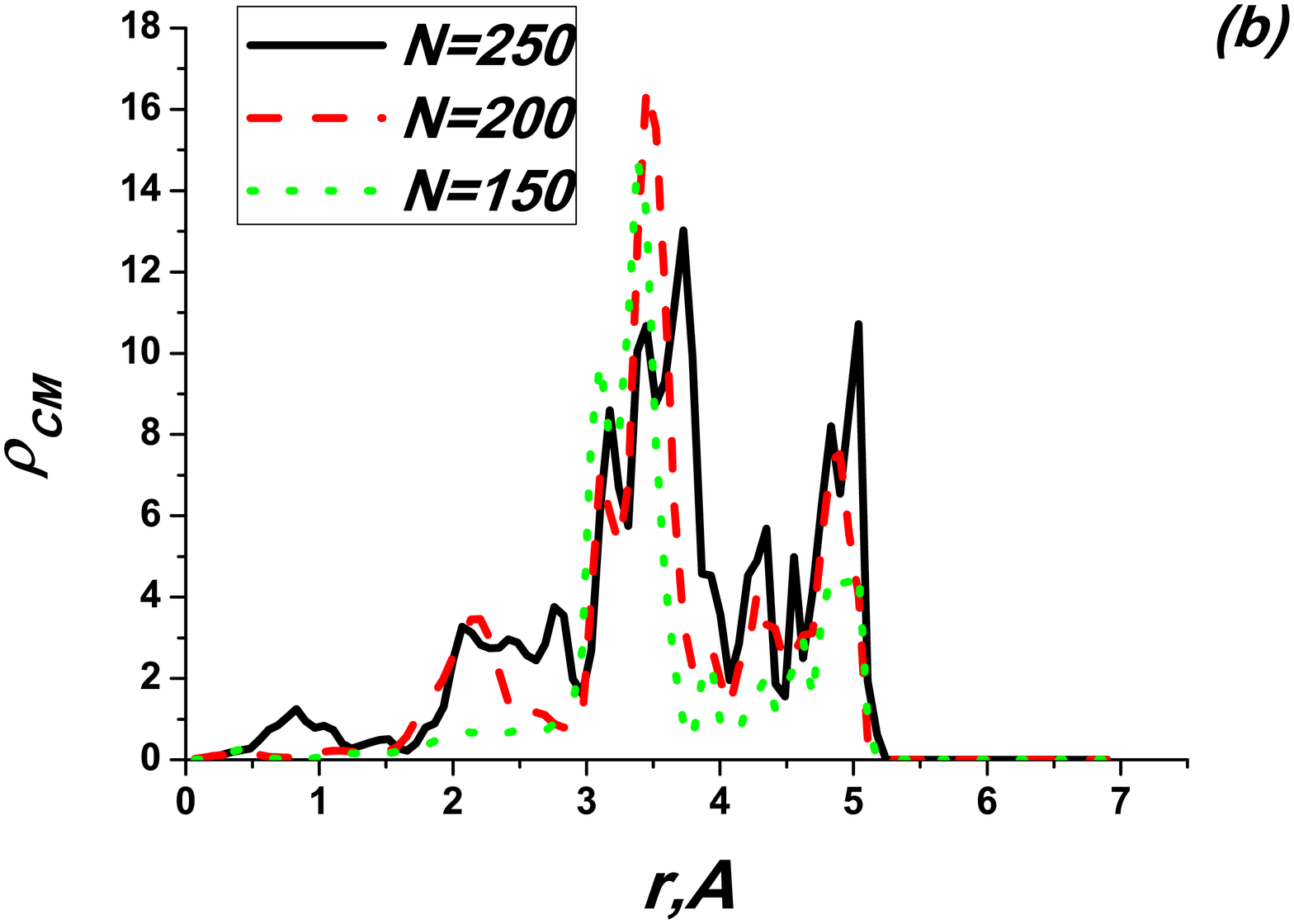}%

\caption{\label{fig:fig3} Radial distributions of number densities
of (a) carbon and (b) centers of mass of the molecules for
different densities of benzene at $T=300K$.}
\end{figure}

Fig.~\ref{fig:fig3} (a) and (b) demonstrate the influence of the
total density $N/V$ on the local densities distribution of carbon
and centers of mass of the molecules. As one expects from general
point of view, as the density increases the peaks become more
pronounced. In the case of the local density of carbon one can see
that at the lower number of molecules ($N=150$ and $200$) the
peaks are almost of the same height. However, as the number of
particles increase the peaks next to the wall rapidly increase
while the increase of the second peak is rather modest.

In the case of the distribution of the centers of mass the
situation is more complex. At the smallest number of molecules
($N=150$) the main peak is located at $3.38 \AA$. As the number of
molecules increases to $N=200$ this peak rises up. However,
further densification of the system leads to placing the molecules
closer to the wall and the second peak ($\approx 5 \AA$) starts to
increase while the first one even decreases with respect to the
previous values.

\begin{figure}
\includegraphics[width=7cm, height=7cm]{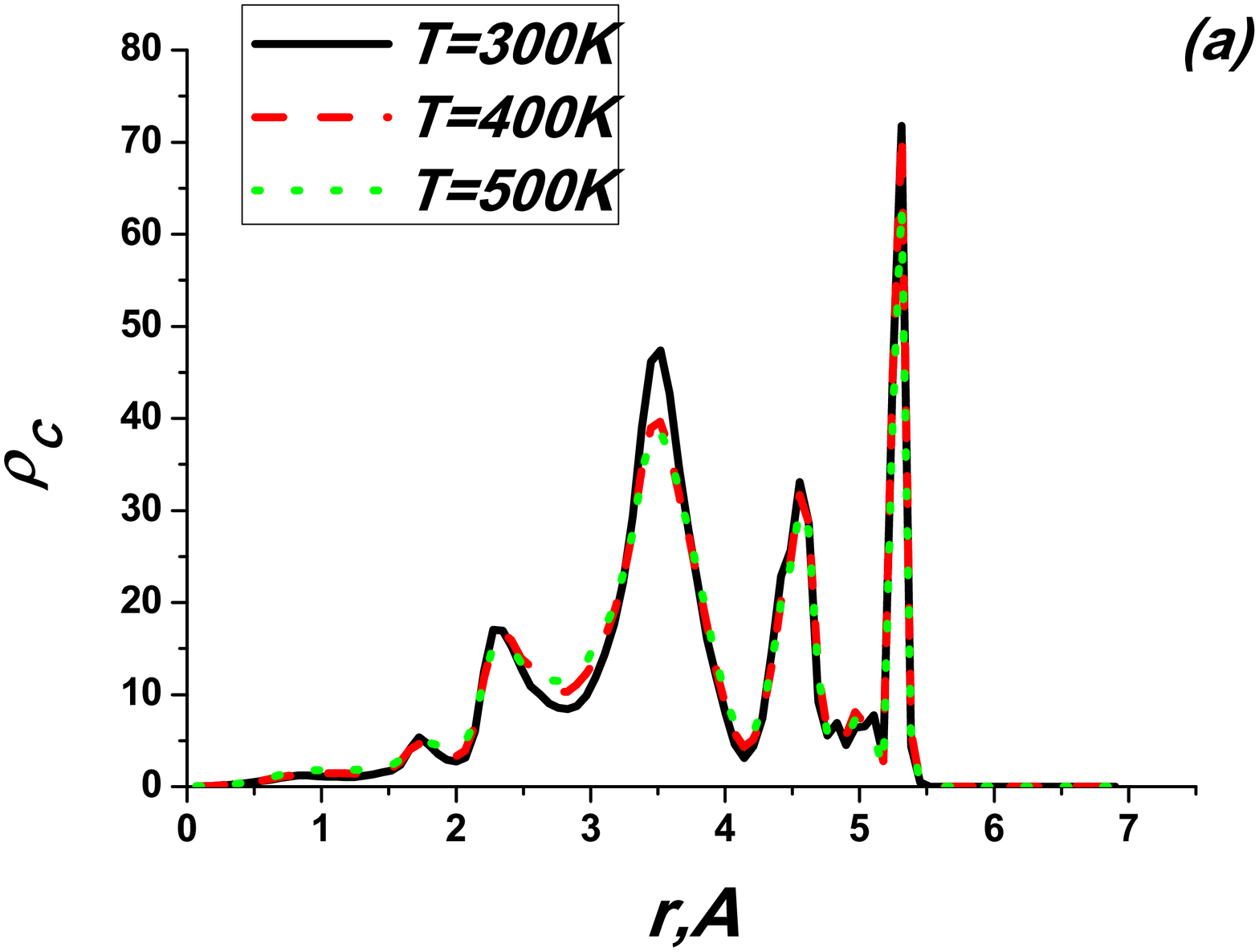}%
\includegraphics[width=7cm, height=7cm]{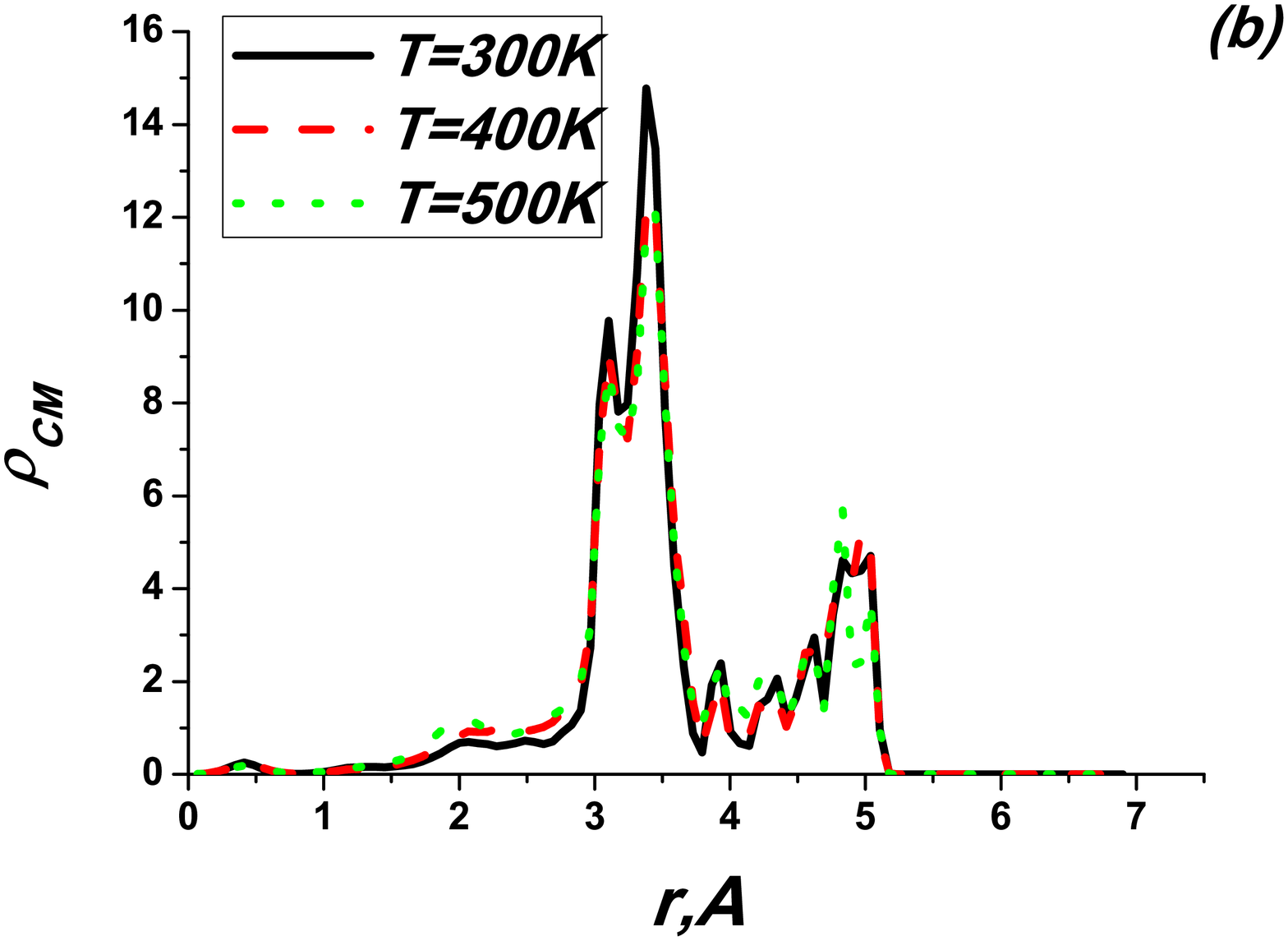}%

\caption{\label{fig:fig4} Radial distributions of number densities
of (a) carbon and (b) centers of mass of the molecules for
different temperatures. The number of benzene molecules inside the
tube is $N=150$.}
\end{figure}

Figs.~\ref{fig:fig4} (a) and (b) demonstrate the influence of
temperature on the density profiles of carbon and centers of mass
of the molecules. As one can expect, the peaks become higher as
the temperature decreases.

In our previous work we studied benzene in graphite and amorphous
carbon slit pores \cite{benzene-jcc}. It was shown that in the
case of graphite walls and relatively small pore sizes (the
distance between the walls below approximately $14.5 \AA$) benzene
molecules form graphite-like sheets. This phenomenon was related
to the close match of the carbon-carbon bond length in graphite
and in benzene ring. Basing on this observation one can ask a
question whether benzene molecules located in the vicinity of the
nanotube mimic the shape of the wall. If so, the molecules loose
theirs planar shape and become scrolled.

In order to estimate the degree of deviation from planar shape of
the benzene ring we employ the following procedure. Denote all
carbon atoms in a ring by numbers from $1$ to $6$ and define the
vectors connecting them: $1-2$, $2-3$, ..., $6-1$. Now we take
three neighboring atoms, say, $1,2,3$. Since they do not belong to
the same straight line there is a unique plane containing these
points. The normal vector of this plane can be identified as a
vector product of the vectors $1-2$ and $1-3$. If we repeat this
procedure for all six carbons of the ring we obtain $6$ vectors
perpendicular to the ring. If the molecule is ideally planar then
all these vectors should be parallel. One can check if these
vectors are parallel by taking theirs scalar product. One can
construct $15$ different pairs. We compute all $15$ scalar
products and sum up the absolute values of the resulting products.
In the case of ideal plane the result should be equal to $15$, so
we divide the final result over $15$. We denote the final quantity
as $P$. It measures the "planarity" of the ring. By definition the
planarity can be less or equal to unity. The deviation of $P$ from
unity can characterize the deviation of benzene molecule from
planar shape.

\begin{figure}
\includegraphics[width=7cm, height=7cm]{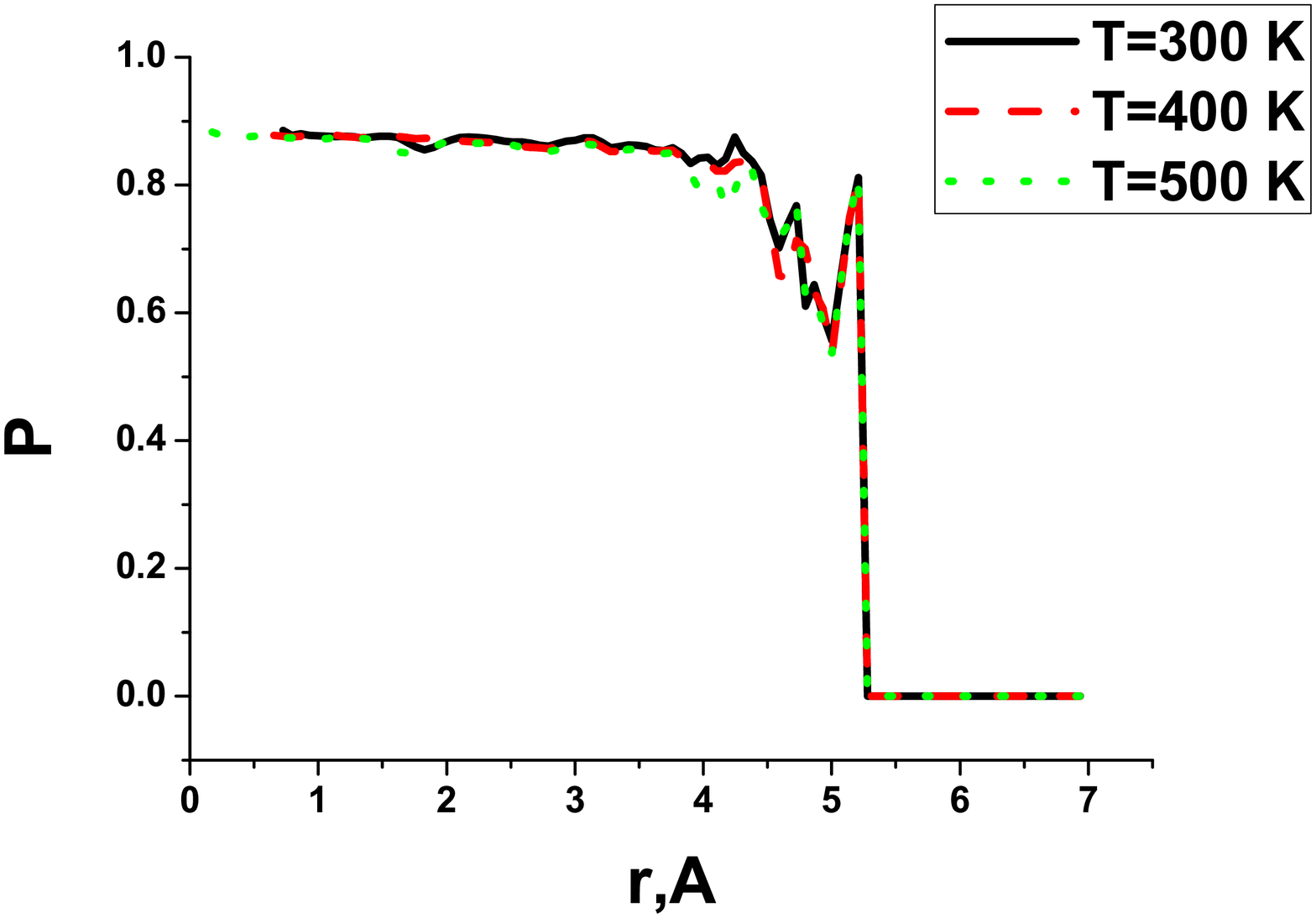}%

\caption{\label{fig:fig5} $P$ parameter which characterizes the
deviation of the benzene ring from planar shape for $N=200$ and
different temperatures.}
\end{figure}

In order to check the validity of the $P$ parameter we calculate
its value for pure bulk benzene at ambient conditions. We find it
to be $P=0.97$ which corresponds to the case of almost planar
rings. In our previous publication we described the structure of
benzene in graphite slit pore \cite{benzene-jcc}. For the case of
the pore size $12.462 \AA$ we find $P=0.95$ which is very close to
the pure benzene result. Fig.~\ref{fig:fig5} shows the $P$
parameter for the case of $N=200$ at three different temperatures.
One can see that up to the distance $r=4 \AA$ the value of $P$ is
$0.87$ which is lower then in the bulk case. Closer to the walls
of the tube the planes become even more distorted. The minimum
value of $P$ is reached at $r=4.8 \AA$ which corresponds to one of
the peaks of the center of mass density distribution. The
magnitude of $P$ at this distance is $P_{min}=0.53$. One can
conclude that close to the walls of the tube the rings bend from
planar shape but the effect is rather weak. The temperature effect
on this planarity parameter is negligible and mostly appears in
low $r$ limit. The curves at different densities also look
qualitatively very similar. For this reason we do not show these
curves for other numbers of particles studied in our work.

Although the deviation of the rings from planar shape can be quite
large one can define the orientational order parameter via second
order Legender polynomial $P_2(cos(\theta))=1.5cos^2(\theta)-0.5$
where $\theta$ is the angle between the normal vector to the plane
of a ring and $z$ direction. If the benzene ring is perpendicular
to the axe of the tube, $P_2=1$ while if the molecule is parallel
to the axe of the tube $P_2=-0.5$. The normal vector to the
benzene ring was defined as an arithmetic average of six vectors
described above for planarity calculations.

\begin{figure}
\includegraphics[width=7cm, height=7cm]{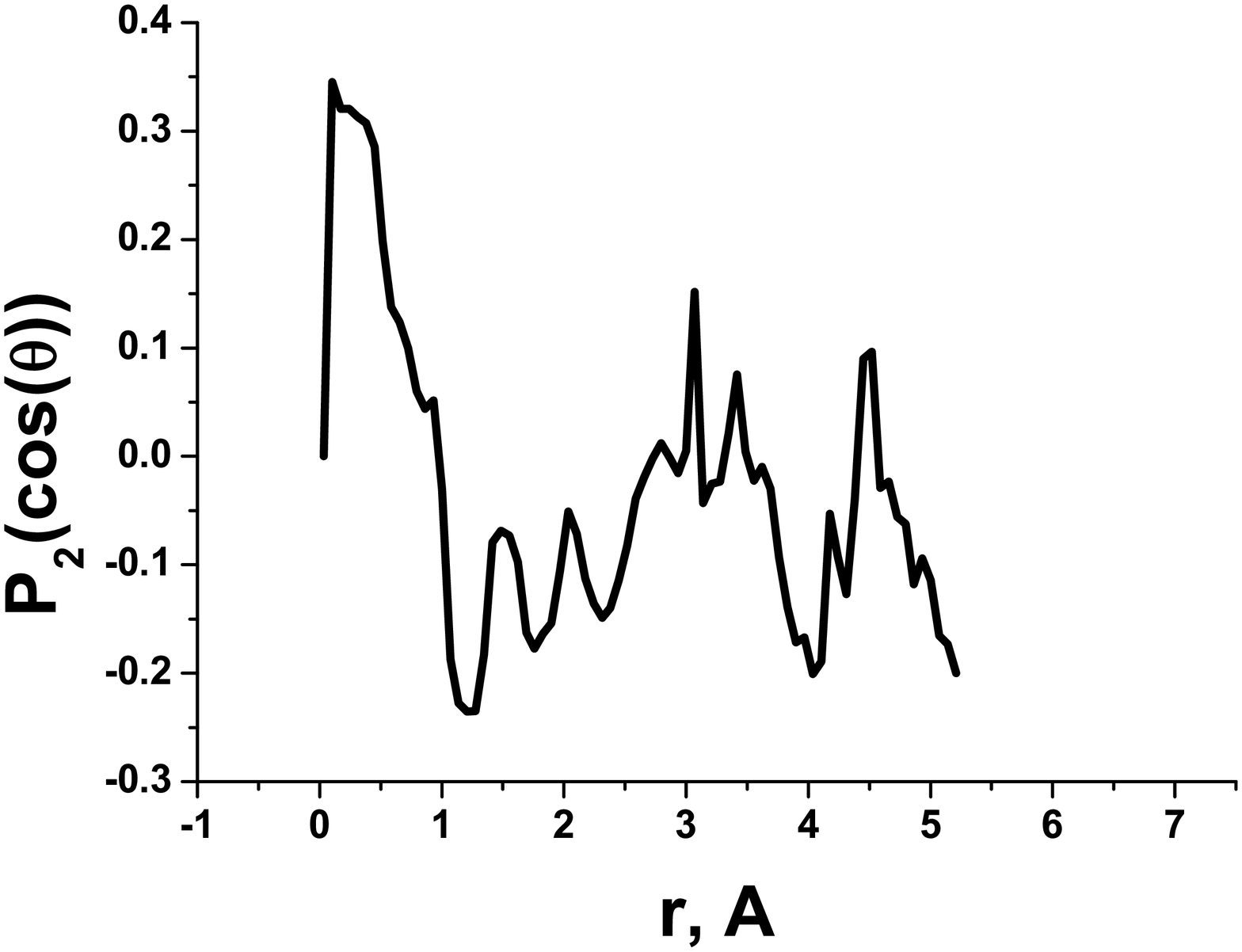}%

\caption{\label{fig:fig6} Legender polynomial $P_2(cos(\theta))$
for $N=250$ and $T=300K$.}
\end{figure}

Fig.~\ref{fig:fig6} shows the radial distribution of $P_2$ for
$N=250$ and $T=300K$, i.e. the highest density and the lowest
temperature studied. One can see that although there is a small
peak at origin the distribution does not demonstrate strong
ordering. It means that unlike the case of graphite slit pores
benzene does not have strong orientational order being confined in
cylindrical geometry. In Ref. \cite{tworings} it was obtained that
for the nanotube radius above $5.580 \AA$ the equilibrium position
of a benzene molecule is the one parallel to the tube axe apart
from the central line. In our simulation we do not find strong
orientational order, however, there is a tilted orientation which
depends on the location of the molecule. Comparison of
Fig.~\ref{fig:fig6} with Fig.~\ref{fig:fig2} shows that inside the
layers where the local density is higher, the benzene molecules
are oriented more perpendicular than between the layers. Mean
field calculations in Ref. \cite{tworings} do not take into
account the heterogenous structure of the system and present only
qualitative tendency in orientations of the benzene molecules.

Finally we discuss the dynamic properties of benzene in carbon
nanotube. Correct calculation of diffusion coefficient requires
simulation in microcanonical ensemble. Our calculations are done
in canonical one. Although such calculations do not give
completely correct numerical value of diffusion coefficient they
allow to get the correct qualitative picture.

\begin{figure}
\includegraphics[width=7cm, height=7cm]{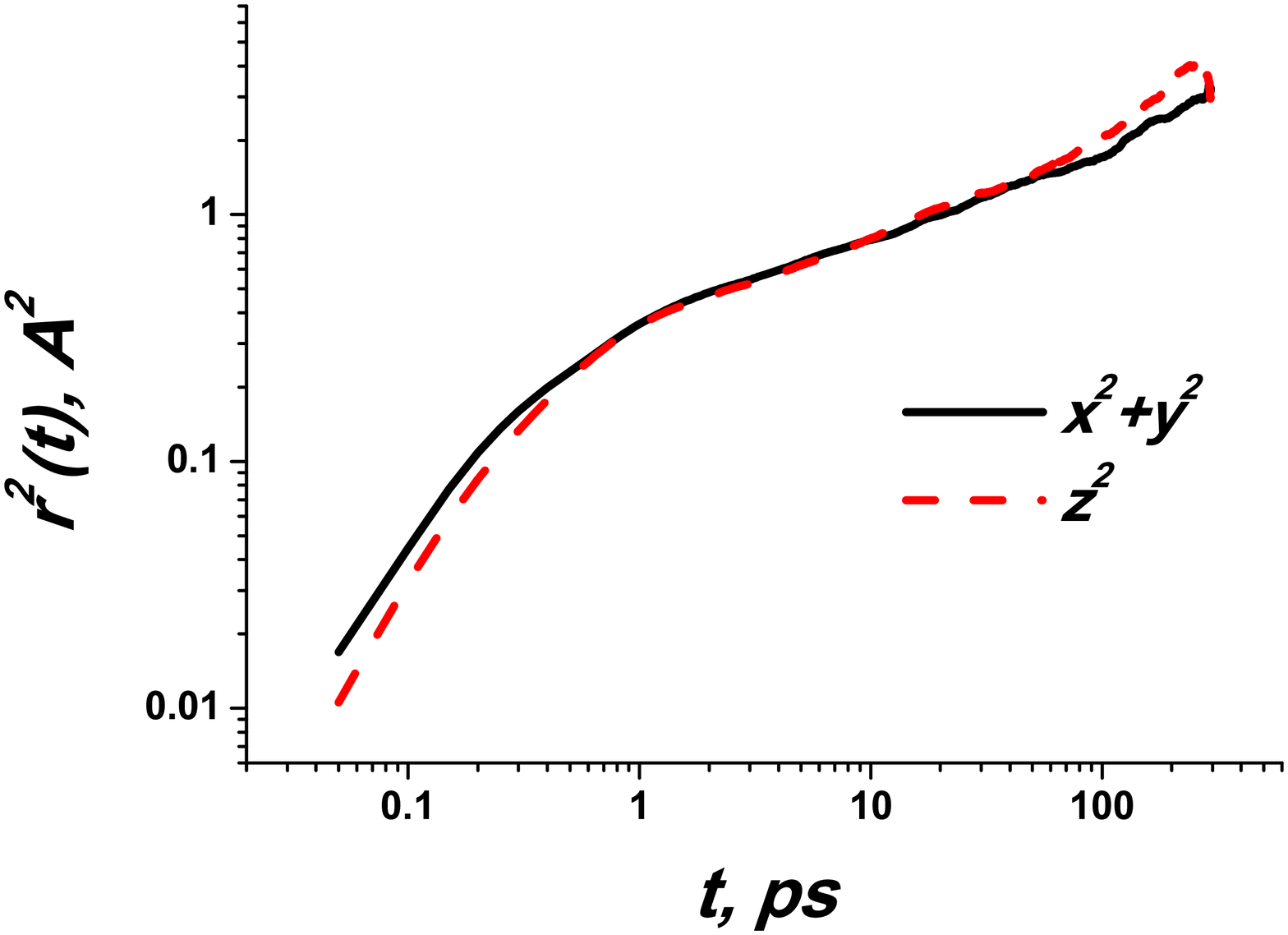}%

\caption{\label{fig:fig7} Mean square displacement of centers of
mass of the molecules in radial direction and in the direction of
the axe of the tube.$N=150$, $T=500 K$}
\end{figure}

The radius of the nanotube we simulate is $6.9 \AA$ which is just
above the size of benzene molecule. It means that the molecules
are confined in a very narrow channel. In such a narrow channel
the system can be roughly considered as one dimensional. In the
case of $1D$ system one does not expect large diffusion. The
particles are strongly caged by theirs nearest neighbors. This is
what we observe in our simulation. Fig.~\ref{fig:fig7} shows the
mean square displacement in the radial directions and in the
direction of the axe of the tube at the lowest density ($N=150$)
and highest temperature $T=500K$ studied. One can see that the
mean square displacement in both directions grows very slowly
which means strong confinement of the molecules. The diffusion
coefficient in both directions is very small in the limits of the
errors of calculations.

\subsection*{\sffamily \large Conclusions}

This paper reports simulation study of benzene molecules confined
into the single wall carbon nanotube. We study the structure and
dynamics of benzene for three densities and three temperatures. We
find that the molecules of benzene mostly group in the middle
distance from the axe of the tube to the wall. The molecules
located in the vicinity of the wall demonstrate some deviation
from planar shape. There is no strong ordering in the system in
contrast to the semianalytical model, proposed in Ref.
\cite{tworings}, however, there is a tilted orientation of the
molecules which depends on the location of the molecule.
Comparison of Fig.~\ref{fig:fig6} with Fig.~\ref{fig:fig2} shows
that inside the layers where the local density is higher, the
benzene molecules are oriented more perpendicular than between the
layers. Last, we find that benzene molecules are almost immobile
at the conditions we report here.

\subsection*{\sffamily \large ACKNOWLEDGMENTS}


Y.F. also thanks the Joint Supercomputing Center of the Russian
Academy of Sciences for computational power and the Russian
Scientific Center Kurchatov Institute for computational
facilities. The work was supported by the Russian Science
Foundation (Grant No 14-12-00820).


\clearpage




\clearpage








\end{document}